\documentclass[3p]{elsarticle}
\usepackage{amsmath}
\usepackage[usenames,dvipsnames]{xcolor}

\usepackage{amssymb}

\usepackage[inline]{enumitem}
\usepackage{hyperref}
\newcommand\currentversion{2019.0}
\newcommand{\MOLSCAT}{{\sc molscat}}
\newcommand{\FIELD}{{\sc field}}
\newcommand{\BOUND}{{\sc bound}}
\newcommand{\etal}{\emph{et al.{}}}
\newcommand\inpitem[1]{{\tt #1}}
\newcommand\basisitem[1]{{\tt #1}}

\newcommand\namelist[1]{{\tt #1}}
\newcommand\code[1]{{\tt #1}}
\newcommand\file[1]{{\tt #1}}
\newcommand\var[1]{{\tt #1}}

\colorlet{crls}{blue}
\colorlet{jmh}{red}

\newcounter{bla}

\journal{Computer Physics Communications}

\begin{document}

\begin{frontmatter}



\title{\BOUND\ and \FIELD: programs for calculating bound states \\
of interacting pairs of atoms and molecules}


\author{Jeremy M. Hutson\corref{jmh}}
\author{C. Ruth Le Sueur\corref{}}

\cortext[jmh]{Corresponding author. \textit{E-mail address:}
J.M.Hutson@durham.ac.uk}
\address{Joint Quantum Centre (JQC)
Durham-Newcastle, Department of Chemistry, \\ University of Durham, South Road,
Durham, DH1 3LE, UK}

\begin{abstract}
The \BOUND\ program calculates the bound states of a complex formed from two
interacting particles using coupled-channel methods. It is particularly
suitable for the bound states of atom-molecule and molecule-molecule van der
Waals complexes and for the near-threshold bound states that are important in
ultracold physics. It uses a basis set for all degrees of freedom except $R$,
the separation of the centres of mass of the two particles. The Schr\"odinger
equation is expressed as a set of coupled equations in $R$. Solutions of the
coupled equations are propagated outwards from the classically forbidden region
at short range and inwards from the classically forbidden region at long range,
and matched at a point in the central region. Built-in coupling cases include
atom + rigid linear molecule, atom + vibrating diatom, atom + rigid symmetric
top, atom + asymmetric or spherical top, rigid diatom + rigid diatom, and rigid
diatom + asymmetric top. Both programs provide an interface for plug-in
routines to specify coupling cases (Hamiltonians and basis sets) that are not
built in. With appropriate plug-in routines, \BOUND\ can take account of the
effects of external electric, magnetic and electromagnetic fields, locating
bound-state energies at fixed values of the fields. The related program \FIELD\
uses the same plug-in routines and locates values of the fields where bound
states exist at a specified energy. As a special case, it can locate values of
the external field where bound states cross scattering thresholds and produce
zero-energy Feshbach resonances. Plug-in routines are supplied to handle the
bound states of a pair of alkali-metal atoms with hyperfine structure in an
applied magnetic field.

\end{abstract}


\end{frontmatter}



{\bf PROGRAM SUMMARIES}

\begin{small}
\noindent
{\em Manuscript Title:}                                       \\
{\em Authors: }Jeremy M. Hutson and C. Ruth Le Sueur                                             \\
{\em Program Titles: } \BOUND\ and \FIELD                                        \\
{\em Journal Reference:}                                      \\
{\em Catalogue identifier:}                                   \\
{\em Licensing provisions:}                                   \\
{\em Programming language: }Fortran 90                                   \\
{\em Computer:}                                               \\
{\em Operating system:}                                       \\
{\em RAM:} case dependent                                              
{\em Classification:} 16.1 Structure and properties             \\
{\em External routines/libraries:} LAPACK, BLAS                           \\
{\em Subprograms used:}                                       \\
{\em Nature of problem:} Solve the Schr\"odinger equation to locate the bound
states of an interacting pair of atoms or molecules as a function of energy
(for \BOUND) or external field (for \FIELD).
\\
   \\
{\em Solution method:}
\\
The Schr\"odinger equation is expressed in terms of coupled equations in the
interparticle distance, $R$. Solutions of the coupled-channel equations are
propagated outwards from the classically forbidden region at short range and
inwards from the classically forbidden region at long range, and matched at a
point in the central region. Bound states exist at energies where one of the
eigenvalues of the log-derivative matching matrix is zero. \BOUND\ calculates
the number of bound states in a specified range of energy and then converges on
the bound-state energies. \FIELD\ operates in a similar manner to \BOUND\ but
converges on bound states as a function of external field at fixed energy, or
energy fixed with respect to a field-dependent threshold energy. The programs
can also generate bound-state wavefunctions if desired.
\\
   \\
   \\
   \\
{\em Restrictions:}\\
   \\
{\em Unusual features: }\\
\begin{enumerate}
\item The programs include Hamiltonians for simple atom-molecule and
    molecule-molecule interactions, and provide an interface that allows
    users to specify Hamiltonians and basis sets for more complex systems.
    This interface allows users to include multiple external fields in the
    Hamiltonian.
\item The programs can propagate very efficiently to long range, making
    them particularly suited to locating very high-lying bound states.
\end{enumerate}
{\em Additional comments:}\\
   \\
{\em Running time: }highly dependent on mass and complexity of interacting particles\\
   \\
\end{small}

\section{Introduction}
\label{intro}

There are many different types of bound-state problem that arise in atomic and
molecular physics. These range from the electronic structure problem, involving
antisymmetrised many-particle wavefunctions and Cou\-lomb interaction
potentials, to low-amplitude molecular vibrational problems that can be solved
in basis sets of harmonic-oscillator functions. In the absence of
wide-amplitude motion, the rovibrational bound-state problem is often
formulated in terms of Eckart-Watson Hamiltonians~\cite{Watson:1968,
Watson:1970, Matyus:2007}. Wide-amplitude motion and exchange of identical
nuclei often require special techniques, even when the motion takes place on a
single electronic potential-energy surface~\cite{Bowman:multimode:2003,
Bowman:variational:2008, McCoy:DMC:2006, Yurchenko:2007}.

The \BOUND\ and \FIELD\ programs deal with an intermediate set of problems
involving interactions between two particles (atoms or molecules), in some
cases on multiple coupled surfaces, where the total Hamiltonian of the system
may be written
\begin{equation}
H=-\frac{\hbar^2}{2\mu}R^{-1}\frac{d^2\,}{d R^2}R
+\frac{\hbar^2 \hat L^2}{2\mu R^2}+H_{\rm intl}(\xi_{\rm intl})+V(R,\xi_{\rm intl}),
\label{eqh}
\end{equation}
where $R$ is a radial coordinate describing the separation of two particles and
$\xi_{\rm intl}$ represents all the other coordinates in the system. $H_{\rm
intl}$ represents the sum of the internal Hamiltonians of the isolated
particles, and depends on $\xi_{\rm intl}$ but not $R$, and $V(R,\xi_{\rm
intl})$ is an interaction potential. The operator $\hbar^2 \hat L^2/2\mu R^2$
is the centrifugal term that describes the end-over-end rotational energy of
the interacting pair.

The Hamiltonian \eqref{eqh} is usually appropriate for pairs of particles that
interact weakly enough that the particles retain their chemical identity. Such
problems commonly arise in the spectroscopy of van der Waals complexes~\cite{Hutson:AMVCD:1991} and in the near-threshold bound states that are
important in the creation and control of ultracold molecules~\cite{Hutson:Cs2:2008}. The internal Hamiltonian $H_{\rm intl}$ is a sum of
terms for the two particles 1 and 2,
\begin{equation}
H_{\rm intl}(\xi_{\rm intl}) = H_{\rm intl}^{(1)}(\xi_{\rm intl}^{(1)})
+ H_{\rm intl}^{(2)}(\xi_{\rm intl}^{(2)}),
\end{equation}
with eigenvalues $E_{{\rm intl},i}=E_{{\rm intl},i}^{(1)}+E_{{\rm
intl},i}^{(2)}$, where $E_{{\rm intl},i}^{(1)}$ and $E_{{\rm intl},i}^{(2)}$
are energies of the separated monomers $1$ and $2$. The individual terms can
vary enormously in complexity: each one may represent a structureless atom,
requiring no internal Hamiltonian at all, a vibrating and/or rotating molecule,
or a particle with electron and/or nuclear spins. The problems that arise in
ultracold physics frequently involve pairs of atoms or molecules with electron
and nuclear spins, often in the presence of external electric, magnetic or
photon fields. All these complications can be taken into account in the
structure of $H_{\rm intl}$ and the interaction potential $V(R,\xi_{\rm
intl})$, which may both involve terms dependent on spins and external fields.

It is possible to solve the bound-state problem for the Hamiltonian \eqref{eqh}
using basis sets for both the internal coordinates $\xi_{\rm intl}$ and the
interparticle distance $R$. Such methods have been used with considerable
success for highly excited states of molecules such as H$_3^+$ and H$_2$O on a
single surface, often using discrete variable
representations~\cite{Tennyson:2004}. However, they have the disadvantage that
the computer time generally scales as the cube of the number of radial basis
functions. This problem becomes worse for levels very close to dissociation. It
can be ameliorated to some extent by using sparse-matrix techniques and
basis-set contraction, but the scaling remains poor.

An alternative is the \emph{coupled-channel} approach, which handles the radial
coordinate $R$ by direct numerical propagation on a grid, and all the other
coordinates using a basis set~\cite{Hutson:CPC:1994}. This is the approach that
is implemented in \BOUND\ and \FIELD. It has the advantage that the computer
time scales \emph{linearly} with the number of points on the radial propagation
grid. In the coupled-channel approach, the total wavefunction is expanded
\begin{equation} \Psi(R,\xi_{\rm intl})
=R^{-1}\sum_j\Phi_j(\xi_{\rm intl})\psi_{j}(R), \label{eqexp}
\end{equation}
where the functions $\Phi_j(\xi_{\rm intl})$ form a complete orthonormal basis
set for motion in the coordinates $\xi_{\rm intl}$ and the factor $R^{-1}$
serves to simplify the form of the radial kinetic energy operator. The
wavefunction in each {\em channel} $j$ is described by a radial \emph{channel
function} $\psi_{j}(R)$. The expansion (\ref{eqexp}) is substituted into the
total Schr\"odinger equation, and the result is projected onto a basis function
$\Phi_i(\xi_{\rm intl})$. The resulting coupled differential equations for the
channel functions $\psi_{i}(R)$ are
\begin{equation}\frac{d^2\psi_{i}}{d R^2}
=\sum_j\left[W_{ij}(R)-{\cal E}\delta_{ij}\right]\psi_{j}(R).
\end{equation}
Here $\delta_{ij}$ is the Kronecker delta and ${\cal E}=2\mu E/\hbar^2$, where
$E$ is the total energy, and
\begin{equation}
W_{ij}(R)=\frac{2\mu}{\hbar^2}\int\Phi_i^*(\xi_{\rm intl}) [\hbar^2 \hat L^2/2\mu R^2 +
H_{\rm intl}+V(R,\xi_{\rm intl})] \Phi_j(\xi_{\rm intl})\,d\xi_{\rm intl}. \label{eqWij}
\end{equation}
The different equations are coupled by the off-diagonal terms $W_{ij}(R)$ with
$i\ne j$.

The coupled equations may be expressed in matrix notation,
\begin{equation}
\frac{d^2\boldsymbol{\psi}}{d R^2}= \left[{\bf W}(R)-{\cal E}{\bf
I}\right]\boldsymbol{\psi}(R). \label{eqcp}
\end{equation}
If there are $N$ basis functions included in the expansion (\ref{eqexp}),
$\boldsymbol{\psi}(R)$ is a column vector of order $N$ with elements
$\psi_{j}(R)$, ${\bf I}$ is the $N\times N$ unit matrix, and ${\bf W}(R)$ is an
$N\times N$ interaction matrix with elements $W_{ij}(R)$.

In general there are $N$ linearly independent solution vectors
$\boldsymbol{\psi}(R)$ that satisfy the Schr\"o\-ding\-er equation subject to
the boundary condition that $\boldsymbol{\psi}(R)\rightarrow0$ at one end of
the range. These $N$ column vectors form a wavefunction matrix
$\boldsymbol{\Psi}(R)$. The propagators in \BOUND\ and \FIELD\ all propagate
the log-derivative matrix ${\bf Y}(R) = \boldsymbol{\Psi}'(R)
[\boldsymbol{\Psi}(R)]^{-1}$, rather than $\boldsymbol{\Psi}(R)$ itself.

The particular choice of the basis functions $\Phi_j(\xi_{\rm intl})$ and the
resulting form of the interaction matrix elements $W_{ij}(R)$ depend on the
physical problem being considered. The complete set of coupled equations often
factorises into blocks determined by the symmetry of the system. In the absence
of external fields, the \emph{total angular momentum} $J_{\rm tot}$ and the
\emph{total parity} are conserved quantities. Different or additional
symmetries arise in different physical situations. The programs are designed to
loop over total angular momentum and parity, constructing a separate set of
coupled equations for each combination and solving them by propagation. These
loops may be repurposed for other symmetries when appropriate.

\BOUND\ and \FIELD\ can also handle interactions that occur in external fields,
where the total angular momentum is no longer a good quantum number.

\subsection{Location of bound states}\label{theory:boundcalcs}

True bound states exist only at energies where all asymptotic channels are
energetically closed, $E<E_{{\rm intl},i}$ for all $i$. Under these
circumstances the bound-state wavefunction $\boldsymbol{\psi}(R)$ is a column
vector of order $N$ that must approach zero in the classically forbidden
regions at both short range, $R\rightarrow 0$, and long range, $R\rightarrow
\infty$.

Continuously differentiable solutions of the coupled equations that satisfy the
boundary conditions at both ends exist only at specific energies $E_n$. These
are the eigenvalues of the total Hamiltonian \eqref{eqh}; we refer to them
(somewhat loosely) as the eigenvalues of the coupled equations, to distinguish
them from eigenvalues of other operators that also enter the discussion below.

Wavefunction matrices $\boldsymbol{\Psi}(R)$ that satisfy the boundary
conditions in \emph{one} of the classically forbidden regions exist at any
energy. We designate these $\boldsymbol{\Psi}^+(R)$ for the solution propagated
outwards from short range and $\boldsymbol{\Psi}^-(R)$ for the solution
propagated inwards from long range. The corresponding log-derivative matrices
are ${\bf Y}^+(R)$ and ${\bf Y}^-(R)$.

It is convenient to choose a matching distance $R_{\rm match}$ where the
outwards and inwards solutions are compared. A solution vector that is
continuous at $R_{\rm match}$ must satisfy
\begin{equation}
\boldsymbol{\psi}(R_{\rm match})=\boldsymbol{\psi}^+(R_{\rm match})=\boldsymbol{\psi}^-(R_{\rm match}).
\end{equation}

Since the derivatives of the outwards and inwards solutions must match too, we
also require that
\begin{equation}
\frac{d}{dR}\boldsymbol{\psi}^+(R_{\rm match})=\frac{d}{dR}\boldsymbol{\psi}^-(R_{\rm match})
\end{equation}
so that
\begin{equation}
{\bf Y}^+(R_{\rm match})\boldsymbol{\psi}(R_{\rm match})
= {\bf Y}^-(R_{\rm match})\boldsymbol{\psi}(R_{\rm match}).
\end{equation}
Equivalently,
\begin{equation}
\left[{\bf Y}^+(R_{\rm match}) - {\bf Y}^-(R_{\rm match})\right]
\boldsymbol{\psi}(R_{\rm match}) = 0,
\label{eq:ymatch}
\end{equation}
so that the wavefunction vector $\boldsymbol{\psi}(R_{\rm match})$ is an
eigenvector of the log-derivative matching matrix, $\Delta{\bf Y} = \left[{\bf
Y}^+(R_{\rm match}) - {\bf Y}^-(R_{\rm match})\right]$, with eigenvalue
zero~\cite{Hutson:CPC:1994}.

For each $J_{\rm tot}$ and symmetry block, \BOUND\ propagates log-derivative
matrices to a matching point $R_{\rm match}$, both outwards from the
classically forbidden region at short range (or from $R=0$) and inwards from
the classically forbidden region at long range. At each energy $E$, it
calculates the multichannel node count~\cite{Johnson:1978}, defined as the
number of zeros of $\boldsymbol{\psi}(R)$ between $R_{\rm min}$ and $R_{\rm
max}$. Johnson \cite{Johnson:1978} showed that this is equal to the number of
eigenvalues of the coupled equations that lie below $E$. It may be calculated
as a simple byproduct of the propagations and the matching matrix. \BOUND\ uses
the node count to determine the number of eigenvalues of the coupled equations
in the specified range, and then uses bisection to identify energy windows that
contain exactly one eigenvalue. In each such window, it uses a combination of
bisection and the Van Wijngaarden-Dekker-Brent algorithm~\cite{VWDB} to
converge on the energy where an eigenvalue of the log-derivative matching
matrix $\Delta {\bf Y}$ is zero. This energy is an eigenvalue of the coupled
equations. The program extracts the local wavefunction vector
$\boldsymbol{\psi}(R_{\rm match})$, and optionally calculates the complete
bound-state wavefunction $\boldsymbol{\psi}(R)$ using the method of Thornley
and Hutson~\cite{THORNLEY:1994}.

\FIELD\ operates in a very similar manner to locate eigenvalues of the coupled
equations as a function of external field at fixed energy (or energy fixed with
respect to a field-dependent threshold energy). The one significant difference
is that the multichannel node count is not guaranteed to be a monotonic
function of field, and it is in principle possible to miss pairs of states that
cross the chosen energy in opposite directions as a function of field. In
practice this seldom happens.

The choice of $R_{\rm match}$ is significant. It does not affect the energies
or fields at which the matching condition (\ref{eq:ymatch}) is satisfied, but
it does affect the matching matrix at other energies or fields and hence the
rate of convergence on eigenvalues of the coupled equations. In particular, it
is usually inappropriate to place $R_{\rm match}$ far into a classically
forbidden region.

\subsection{Matrix of the interaction potential}\label{theory:W}

In order to streamline the calculation of matrix elements for the propagation,
\BOUND\ and \FIELD\ express the interaction potential in an expansion over the
internal coordinates,
\begin{equation}
V(R,\xi_{\rm intl})=\sum_\Lambda v_\Lambda(R){\cal V}^\Lambda(\xi_{\rm intl}).
\label{eqvlambda}
\end{equation}
The specific form of the expansion depends on the nature of the interacting
particles. The radial potential coefficients $v_\Lambda(R)$ may either be
supplied explicitly, or generated internally by numerically integrating over
$\xi_{\rm intl}$. The $R$-independent coupling matrices $\boldsymbol{\cal
V}^\Lambda$ with elements ${\cal V}^\Lambda_{ij}=\langle\Phi_i|{\cal
V}^\Lambda|\Phi_j\rangle_{\rm intl}$ are calculated once and stored for use in
evaluating $W_{ij}(R)$ throughout the course of a propagation.

\subsection{Matrices of the internal and centrifugal Hamiltonians}\label{theory:Wextra}

Coupled-channel theory is most commonly formulated in a basis set where $\hat
L^2$ and $H_{\rm intl}$ are both diagonal. All the built-in coupling cases use
basis sets of this type. The matrix of $H_{\rm intl}$ is $\langle\Phi_i|H_{\rm
intl}|\Phi_j\rangle_{\rm intl}=E_{{\rm intl},i}\delta_{ij}$. The diagonal
matrix elements of $\hat L^2$ are often of the form $L_i(L_i+1)$, where the
integer quantum number $L_i$ (sometimes called the partial-wave quantum number)
represents the end-over-end angular momentum of the two particles about one
another.

However, the programs also allow the use of basis sets where one or both of
$\hat L^2$ and $H_{\rm intl}$ are non-diagonal. If $H_{\rm intl}$ is
non-diagonal, it is expanded as a sum of terms
\begin{equation}
H_{\rm intl}(\xi_{\rm intl})
=\sum_\Omega h_\Omega {\cal H}^\Omega_{\rm intl}(\xi_{\rm intl}),
\label{eqHomega1}
\end{equation}
where the $h_\Omega$ are scalar quantities, some of which may represent
external fields if desired. The programs generate additional coupling matrices
$\boldsymbol{\cal H}^\Omega$ with elements ${\cal
H}^\Omega_{ij}=\langle\Phi_i|{\cal H}^\Omega_{\rm intl}|\Phi_j\rangle_{\rm
intl}.$ These are also calculated once and stored for use in evaluating
$W_{ij}(R)$ throughout the course of a propagation. A similar mechanism is used
for basis sets where $\hat L^2$ is non-diagonal, with
\begin{equation}
\hat L^2
=\sum_\Upsilon {\cal L}^\Upsilon.
\label{eqL2}
\end{equation}

If $H_{\rm intl}$ is non-diagonal, the allowed energies $E_{{\rm intl},i}$ of
the pair of monomers at infinite separation are the eigenvalues of $H_{\rm
intl}$. The wavefunctions of the separated pair are represented by simultaneous
eigenvectors of $H_{\rm intl}$ and $\hat L^2$.

\subsection{Boundary conditions}

For deeply bound states, it is often sufficient to require that
$\boldsymbol{\Psi}(R)\rightarrow0$ in the classically forbidden regions at
short and long range, or equivalently that ${\bf Y}(R)\rightarrow\pm\infty$.
However, there are circumstances where more general boundary conditions are
required:
\begin{itemize}[nosep]
\item In systems where $R=0$ is energetically accessible, some states
    require ${\bf Y}(0)=0$.
\item In model systems with a Fermi pseudopotential, corresponding to a
    $\delta$-function at the origin or elsewhere, a finite value of ${\bf
    Y}$ may be required.
\item For states very close to dissociation, the wavefunction
    $\boldsymbol{\psi}(R)$ dies off very slowly at long range, and it may
    be inefficient to propagate far enough that $\boldsymbol{\psi}(R)
    \rightarrow 0$. For a single channel, the wavefunction approximately
    follows the Wentzel-Kramers-Brillouin (WKB) approximation in the far
    classically forbidden region,
\begin{eqnarray}
\psi(R)&=&[k(R)]^{-\frac{1}{2}} \exp\left(\pm\int_{R_{\rm turn}}^R k(R')\,d R'\right),\\
\psi'(R)&=&[k(R)]^{-\frac{1}{2}}\left[\pm k(R)-\frac{1}{2}\frac{k'(R)}{k(R)} \right]
\exp\left(\pm\int_{R_{\rm turn}}^R k(R')\,d R'\right),\\
Y(R)&=&\pm k(R)-\frac{1}{2}\frac{k'(R)}{k(R)},\label{eq:bcwkb}
\end{eqnarray}
where $k(R) = [2\mu(V(R)-E)/\hbar^2]^{1/2}$ and $V(R)$ is an effective
potential energy for the channel concerned. The $+$ sign applies inside the
inner turning point (where the phase integral itself is negative) and the
$-$ sign applies outside the outer turning point. The first term in Eq.\
\ref{eq:bcwkb} dominates either when $k(R)$ is large (in a strongly
classically forbidden region) or when the interaction potential is nearly
constant (at very long range). The term involving $k'(R)$ is therefore
neglected in the implementation of WKB boundary conditions.
\end{itemize}
\BOUND\ and \FIELD\ allow the imposition of separate boundary conditions for
${\bf Y}$ in closed and in open channels at $R_{\rm min}$ and at $R_{\rm max}$, and
by default apply WKB boundary conditions for closed channels (neglecting the
term involving $k'(R)$ in Eq.\ \ref{eq:bcwkb}). This gives faster convergence
with respect to $R_{\rm min}$ and $R_{\rm max}$ than ${\bf
Y}(R)\rightarrow\pm\infty$.

\subsection{Perturbation calculations}

\BOUND\ can calculate expectation values using a finite-difference approach
\cite{Hutson:expect:88}. After a bound state is located at energy $E_n^{(0)}$,
\BOUND\ repeats the calculation with a small perturbation $a \hat A(R)$ added
to the Hamiltonian to obtain a modified energy $E_n(a)$. From perturbation
theory,
\begin{equation} E_n(a) = E_n^{(0)} + a
\langle\hat A\rangle_n + {\cal O}(a^2),
\end{equation}
where ${\cal O}(a^2)$ are second-order terms. The finite-difference
approximation to the expectation value $\langle\hat A\rangle_n$ is
\begin{equation}
\langle\hat A\rangle_n = \frac{E_n(a) - E_n^{(0)}}{a},
\end{equation}
and is accurate to order $a$.

For built-in coupling cases, \BOUND\ can calculate expectation values of an
operator $\hat A$ that is made up of a product of one of the angular functions
in the potential expansion and a power of $R$. For coupling cases implemented
in plug-in basis-set suites, any required operator can be implemented in the
basis-set suite.

\subsection{Richardson extrapolation}

For propagators that use equally spaced steps, the error in bound-state
energies due to a finite step size is proportional to a power of the step size
(in the limit of small steps). \BOUND\ can obtain an
improved estimate of the bound-state energy by performing calculations with two
different step sizes and extrapolating to zero step size.

\subsection{Reference energies}

By default, the zero of energy used for total energies is the one used for
monomer energies, or defined by the monomer Hamiltonians programmed in a
plug-in basis-set suite. However, it is sometimes desirable to use a different
zero of energy (reference energy). This may be specified:
\begin{itemize}[nosep]
\item{as a value given directly in the input file;}
\item{as the energy of a particular scattering threshold or pair of monomer
    states, which may depend on external fields.}
\end{itemize}

\subsection{Locating zero-energy Feshbach resonances}

Zero-energy Feshbach resonances occur at fields where bound states cross a
scattering threshold as a function of external field, provided there is
coupling between the bound state and the threshold. \FIELD\ may be used to
locate such crossings by choosing the energy of the desired threshold as the
reference energy and setting the relative energy to zero.

\subsection{Wavefunctions}

The programs always extract the local wavefunction vector
$\boldsymbol{\psi}(R_{\rm match})$ at the matching point. If desired, they can
calculate the complete bound-state wavefunction $\boldsymbol{\psi}(R)$ using
the method of Thornley and Hutson~\cite{THORNLEY:1994}.

\section{Systems handled}\label{interactiontypes}

The programs provide built-in Hamiltonians and basis sets to handle a number of
common cases. In particular, they can calculate bound states in the
close-coupling approximation (with no dynamical approximations except basis-set
truncation) for the following pairs of species:
\begin{enumerate}[nosep]
\item Atom + linear rigid rotor~\cite{Arthurs:1960};
\item Atom + vibrating diatom with interaction potentials independent of
    diatom rotational state~\cite{Green:1979:vibrational};
\item Linear rigid rotor + linear rigid
    rotor~\cite{Green:1975,Green:1977:comment,Heil:1978:coupled};
\item Asymmetric top + linear molecule~\cite{Phillips:1995}
\item Atom + symmetric top (also handles near-symmetric tops and linear
    molecules with vibrational angular
    momentum)~\cite{Green:1976,Green:1979:IOS};
\item Atom + asymmetric top~\cite{Green:1976} (also handles spherical
    tops~\cite{Hutson:spher:1994});
\item Atom + vibrating diatom with interaction potentials dependent on
    diatom rotational state~\cite{Hutson:sbe:1984};
\item Atom + rigid corrugated surface
    \cite{Wolken:1973:surface,Hutson:1983} (band structure). At present,
    the code is restricted to centrosymmetric lattices, for which the
    potential matrices are real, and is not included in \FIELD.
\end{enumerate}
The close-coupling calculations are all implemented in a fully coupled
space-fixed representation, with the calculations performed separately for each
total angular momentum and parity.

In addition, the programs implement a variety of dynamical approximations
(decoupling methods) that offer considerable savings of computer time at the
expense of accuracy. Some of these are of significance only for scattering
calculations, such as the effective potential approximation~\cite{Rabitz:EP},
the $L$-labelled coupled-states (CS) approximation~\cite{McG74} and the
decoupled $L$-dominant approximation~\cite{Green:1976:DLD,DePristo:1976:DLD}.
However, bound-state calculations frequently use the helicity decoupling
approximation~\cite{Hutson:AMVCD:1991}, which is implemented in \BOUND\ and
\FIELD\ in the framework of the CS approximation.

In addition to the built-in cases, the programs provide an interface that
allows users to specify Hamiltonians and basis sets for different pairs of
species. These have been used for numerous different cases, and routines are
provided for two cases of current interest:
\begin{enumerate}[nosep]
\item Structureless atom + $^3\Sigma$ molecule in a magnetic field,
    demonstrated for Mg + NH;
\item Two alkali-metal atoms in $^2$S states, including hyperfine coupling,
    in a magnetic field, demonstrated for $^{85}$Rb$_2$.
\end{enumerate}

\section{Propagators}\label{propagators}

\BOUND\ and \FIELD\ can solve the coupled equations using a variety of
different propagation methods, all of which propagate the log-derivative matrix
${\bf Y}(R)$ rather than the wavefunction matrix $\boldsymbol{\Psi}(R)$.
Log-derivative propagators are particularly suitable for the bound-state
problem, both because they allow a very simple form of the matching equation
and because they inherently avoid the instability associated with propagating
in the presence of deeply closed channels. The propagators currently
implemented in \BOUND\ and \FIELD\ are:
\begin{itemize}[leftmargin=13pt]
\item{Log-derivative propagator of Johnson (LDJ)~\cite{Johnson:1973,
    Manolopoulos:1993:Johnson}: This is a very stable propagator. It has
    largely been superseded by the LDMD propagator, but can be useful in
    occasional cases where that propagator has trouble evaluating node
    counts.}
\item{Diabatic log-derivative propagator of Manolopoulos
    (LDMD)~\cite{Manolopoulos:1986}: This is a very efficient and stable
    propagator, especially at short and medium range.}
\item{Quasiadiabatic log-derivative propagator of Manolopoulos
    (LDMA)~\cite{Manolopoulos:PhD:1988, Hutson:CPC:1994}: This is similar
    to the LDMD propagator, but operates in a quasiadiabatic basis. It
    offers better accuracy than LDMD for very strongly coupled problems,
    but is relatively expensive. It is recommended for production runs only
    for very strongly coupled problems. However, it is also useful when
    setting up a new system, because it can output eigenvalues of the
    interaction matrix at specific distances (adiabats) and nonadiabatic
    couplings between the adiabatic states.}
\item{Symplectic log-derivative propagators of Manolopoulos and Gray (LDMG)
    \cite{MG:symplectic:1995}: This offers a choice of 4th-order or
    5th-order symplectic propagators. These are 1.5 to 3 times more
    expensive per step than the LDMD and LDJ propagators, but can have
    smaller errors for a given step size.  They can be the most efficient
    choice when high precision is required.}
\item{Airy propagator: This is the AIRY log-derivative propagator of
    Alexander~\cite{Alexander:1984} as reformulated by Alexander and
    Manolopoulos~\cite{Alexander:1987}.  It uses a quasiadiabatic basis
    with a linear reference potential (which results in Airy functions as
    reference solutions). This allows the step size to increase rapidly
    with separation, so that this propagator is particularly efficient at
    long range.}
\end{itemize}

Calculations with \BOUND\ and \FIELD\ may use different log-derivative
propagators at short and long range. This is particularly useful for bound
states close to a dissociation threshold, where it may be necessary to
propagate inwards from very large values of $R$ to obtain converged results.
The AIRY propagator incorporates a variable step size, and can be used to
propagate inwards with an initially very large but decreasing step size at very
low cost. However, it is not particularly efficient when the interaction
potential is fast-varying, so it is often used in combination with a
fixed-step-size method such as the LDMD propagator at short and intermediate
range.

\section{Computer time}

The computer time required to solve a set of $N$ coupled equations is
approximately proportional to $N^3$. The practical limit on $N$ is from a few
hundred to several thousand, depending on the speed of the computer and the
amount of memory available.

The computer time also depends linearly on the number of radial steps required
to solve the coupled equations to the desired accuracy. The step size required
is typically proportional to the minimum local wavelength, so that the time
scales approximately with $(\mu E_{\rm max}^{\rm kin})^{1/2}$, where $E_{\rm
max}^{\rm kin}$ is the maximum local kinetic energy; for bound states near
dissociation, $E_{\rm max}^{\rm kin}$ may be approximated by the well depth of
the interaction potential.

\section{Plug-in functionality}

\subsection{Potential or potential expansion coefficients}

The programs internally express the interaction potential as an expansion over
the internal coordinates, as in Eq.~\eqref{eqvlambda}. The expansion
coefficients $v_\Lambda(R)$ may be supplied in a variety of ways:
\begin{itemize}[nosep, leftmargin=13pt]
\item For very simple potentials, where the functions $v_\Lambda(R)$ are
    sums of exponentials and inverse powers, the parameters that specify
    them may be supplied in the input file.
\item For more complicated functions, plug-in routines may be supplied to
    return individual values of $v_\Lambda(R)$ at a value of $R$ specified
    in the calling sequence.
\item For most of the built-in coupling cases, plug-in routines may be
    supplied to return the unexpanded potential $V(R,\xi_{\rm intl})$ at
    specified values of $R$ and the internal coordinates $\xi_{\rm intl}$.
    The general-purpose potential routine supplied then performs numerical
    quadrature over $\xi_{\rm intl}$ to evaluate the expansion coefficients
    $v_\Lambda(R)$ internally.
\item If none of these approaches is convenient (or efficient enough), a
    replacement potential routine may be supplied to return the complete
    potential expansion at once (all values of $v_\Lambda(R)$ at a value of
    $R$ specified in the calling sequence).
\end{itemize}

\subsection{Basis sets and coupling matrices}

The programs provide an interface for users to supply a set of routines that
specify an additional type of basis set, select the elements that will be used
in a particular calculation, and calculate the matrices of coupling
coefficients for the operators ${\cal V}^\Lambda(\xi_{\rm intl})$ used to
expand the interaction potential. The routines must also specify the matrices
of $H_{\rm intl}$ and $\hat L^2$, which may be diagonal or non-diagonal. If
desired, $H_{\rm intl}$ may contain terms that depend on external fields.

\subsection{External fields and potential scaling}

The programs incorporate data structures to handle external electric, magnetic
or photon fields. There may be multiple fields at arbitrary relative
orientations. Internally, the field-dependent terms in the Hamiltonian are a
subset of those in $H_{\rm intl}$,
\begin{equation}
H_{\rm intl}(\xi_{\rm intl},\boldsymbol{B})
=\sum_\Omega B_\Omega {\cal H}^\Omega_{\rm intl}(\xi_{\rm intl}),
\label{eqHomegaB}
\end{equation}
where the vector $\boldsymbol{B}$ represents all the fields present. The
elements of $\boldsymbol{B}$ may each be expressed as a \emph{nonlinear}
function of external field variables (EFVs); the EFVs may thus (for example)
represent the magnitudes, orientations, or relative angles of the component
fields. \BOUND\ allows calculations on a grid of values of any one EFV, and
\FIELD\ allows bound states to be located as a function of one EFV with all the
others fixed.

The programs also allow calculations as a function of a scaling factor that
multiplies the entire interaction potential, or a subset of the potential
expansion coefficients $v_\Lambda(R)$. The scaling factor is handled internally
using the same structures as external fields.

\section{Distributed files and example calculations}\label{use:input}

\subsection{Distributed files}

The program is supplied as a tarred zipped file, which contains:
\begin{itemize}
\item{the full program documentation in pdf format;}
\item{a directory {\tt source\_code} containing
\begin{itemize}
\item{the Fortran source code;}
\item{a GNU makefile ({\tt GNUmakefile}) that can build the executables
    needed for the example calculations;}
\end{itemize}}
\item{a directory {\tt examples} containing
\begin{itemize}
\item{a sub-directory {\tt input} containing input files for the
    example calculations described below;}
\item{a sub-directory {\tt output} containing the corresponding output files;}
\end{itemize}}
\item{a directory {\tt data} containing auxiliary data files
    for some potential routines used in the example calculations;}
\item{a plain-text file {\tt README} that gives information on changes that
    may be needed to adapt the GNUmakefile to a specific target computer.}
\item{a plain-text file {\tt COPYING} that contains the text of the GNU General Public License, Version 3.}
\end{itemize}

\subsection{Example calculations}

The executables used for different calculations may differ in the routines
linked to construct the basis set, specify the internal Hamiltonian, and
evaluate the interaction potential. The executables required for the example
calculations can all be built using {\tt GNUmakefile}.

\subsubsection{All available propagators}\label{testfiles:bound:intflgs}

\begin{tabular}{ll}
input file: & \file{bound-all\_propagators.input}\\
executable: & \file{bound-basic}
\end{tabular}

\file{bound-all\_propagators.input} performs close-coupling calculations on the
bound states of a simple model of a complex formed between an atom and a linear
rigid rotor. The radial potential coefficients are provided in the input data
file and consist of a Lennard-Jones 12-6 potential for $\lambda=0$ and a
dispersion-like $R^{-6}$ form for $\lambda=2$. The calculation is repeated
using combinations of short-range and long-range propagators that exercise
every propagation method available in \BOUND\ (though not every possible
combination). The calculation is done twice for the LDMD/AIRY combination; once
with $R_{\rm mid} < R_{\rm match}$ and once with $R_{\rm mid}
> R_{\rm match}$. The calculation which uses just the LDMD propagator employs a
different step length for the inwards propagation. This input file should
produce the same results regardless of which \BOUND\ executable is used.

\subsubsection{Bound states of Ar-HCl with expectation
values}\label{testfiles:bound:ityp1}

\begin{tabular}{ll}
input file: & \file{bound-Ar\_HCl.input}\\
executable: & \file{bound-Rg\_HX}
\end{tabular}

\file{bound-Ar\_HCl.input} performs calculations on the states of Ar-HCl bound
by more than 80 cm$^{-1}$, using the H6(4,3,0) potential of Hutson~\cite{H92ArHCl}
and the LDMD propagator, for total angular momentum $J_{\rm tot}=0$ and 1
and both parities. The first run does close-coupling calculations. The second
run does calculations in the helicity decoupling approximation, and in addition
calculates expectation values $\langle P_2(\cos\theta)\rangle$ and $\langle
1/R^2 \rangle$ for all the states. The results may be compared with Table IV of
ref.~\cite{H92ArHCl}.  The third run calculates the wavefunction for the first bound state identified in the first run. The wavefunction is written to unit 109; the resulting file is included as \file{bound-Ar\_HCl.wavefunction} in \file{examples/output}. The components may be plotted with any standard plotting package.

\subsubsection{\texorpdfstring{Bound states of Ar-CO$_2$ with Richardson extrapolation}{Bound states of Ar-CO2 with Richardson extrapolation}}\label{testfiles:bound:ityp1b}

\begin{tabular}{ll}
input file: & \file{bound-Ar\_CO2.input}\\
executable: & \file{bound-Rg\_CO2}
\end{tabular}

\file{bound-Ar\_CO2.input} performs close-coupling calculations on the ground
and first vibrationally excited state of Ar-CO$_2$, using the split repulsion
potential of Hutson \etal~\cite{H96ArCO2fit} and the LDJ propagator, for total
angular momentum $J_{\rm tot}=0$. The results may be compared with Table IV of
ref.~\cite{H96ArCO2fit}.

It first calculates the ground-state energy using a fairly large (unconverged)
step size of 0.03 \AA. It then repeats the calculation with an even larger step
size, and extrapolates to zero step size using Richardson $h^4$ extrapolation.

\subsubsection{\texorpdfstring{Bound states of Ar-H$_2$}{Bound states of Ar-H2}}\label{testfiles:bound:ityp7}

\begin{tabular}{ll}
input file: & \file{bound-Ar\_H2.input}\\
executable: & \file{bound-Rg\_H2}\\
also required: & \file{data/h2even.dat}
\end{tabular}

\file{bound-Ar\_H2.input} performs close-coupling calculations on the ground
state of Ar-H$_2$ with H$_2$ in its $v=1$, $j=1$ state, for total angular
momentum $J_{\rm tot}=1$ and even parity ($j+L$ even). For this parity there is no
allowed $j=0$ channel, so the state is bound except for vibrational
predissociation to form H$_2$ ($v=0$)~\cite{HUTSON:ArH2:1983}, which is not
taken into account by \BOUND. The run uses the LDMD propagator and the TT3(6,8)
potential of Le~Roy and Hutson~\cite{LeR87}, evaluated for H$_2$ states $(j,v)
= (0,0)$, (2,0) and (4,0) using H$_2$ matrix elements provided in the file
\file{data/h2even.dat}.

\BOUND\ first calculates the ground-state energy using a fairly large
(unconverged) step size of 0.04 \AA. It then repeats the calculation with an
even larger step size, and extrapolates to zero step size using Richardson
$h^4$ extrapolation.

\subsubsection{\texorpdfstring{Bound states of H$_2$-H$_2$ (ortho-para)}{Bound states of H2-H2 (ortho-para)}}\label{testfiles:bound:ityp3}

\begin{tabular}{ll}
input file: & \file{bound-ityp3.input}\\
executable: & \file{bound-H2\_H2}
\end{tabular}

\file{bound-ityp3.input} performs close-coupling calculations on bound states
of H$_2$-H$_2$ with one para-H$_2$ molecule (even $j$) and one ortho-H$_2$
molecule (odd $j$). It uses the LDMD propagator. The interaction potential is
that of Zarur and Rabitz~\cite{Zarur:1974}. The states are bound by less than
2~cm$^{-1}$ (below the $j=0$ + $j=1$ threshold).

\subsubsection{\texorpdfstring{Bound states of He-NH$_3$}{Bound states of He-NH3}}\label{testfiles:bound:ityp5}

\begin{tabular}{ll}
input file: & \file{bound-ityp5.input}\\
executable: & \file{bound-basic}
\end{tabular}

\file{bound-ityp5.input} performs close-coupling calculations on bound states
of He-NH$_3$, taking account of the tunnelling splitting of NH$_3$, using a
simple analytical interaction potential and the LDMD propagator. The input file
selects rotational functions of E symmetry by setting \basisitem{ISYM(3)} to 1
and specifies that the H nuclei are fermions by setting \basisitem{ISYM(4)} to
1.

\subsubsection{\texorpdfstring{Bound states of Ar-CH$_4$}{Bound states of Ar-CH4}}\label{testfiles:bound:ityp6}

\begin{tabular}{ll}
input file: & \file{bound-Ar\_CH4.input}\\
executable: & \file{bound-Ar\_CH4}
\end{tabular}

\file{bound-Ar\_CH4.input} performs close-coupling calculations on bound states
of Ar-CH$_4$, using $\basisitem{ITYPE}=6$. It uses the interaction potential of Buck
\etal~\cite{Buck:1983}. It uses the LDMD propagator. CH$_4$ is a spherical top,
and the input file selects rotor functions of F symmetry and even $k$ by
setting \basisitem{ISYM} to 177. The results may be compared with Table II of
ref.~\cite{Hutson:spher:1994}.

%
%

\subsubsection{Bound-state energies of the hydrogen atom}\label{testfiles:bound:H}

\begin{tabular}{ll}
input file: & \file{bound-hydrogen.input}\\
executable: & \file{bound-basic}
\end{tabular}

\file{bound-hydrogen.input} carries out single-channel bound-state calculations
on the hydrogen atom, and demonstrates how to handle calculations in atomic
units. It sets \inpitem{MUNIT} to the electron mass in Daltons, \inpitem{RUNIT}
to the Bohr radius in \AA\ and $\inpitem{EUNITS}=7$ to select input energies in
hartrees. It sets up a simple Coulomb potential, with the energy scaling factor
set to the hartree in cm$^{-1}$, so that the potential is handled in atomic
units. It uses the atom-rigid rotor basis with $\basisitem{JMAX}=0$ to generate
a simple single-channel problem. Note that \basisitem{ROTI}\code{(1)} is set to
the dummy value \code{1.0}; this value is not used because
$\basisitem{JMAX}=0$, but it prevents the program terminating prematurely.

The wavefunction at the origin is of the form $r^{l+1}$, so its log-derivative
is infinite at the origin. This is the default for locally closed channels,
but is specified explicitly for the locally open $l=0$ channel.

Because $\basisitem{JMAX}=0$, the orbital angular momentum $l$ is equal to
\var{JTOT}. $\var{JTOT}=0$ produces $n$s levels at energies of $-1/(2n^2)$ for
$n=1,2,...$, while $\var{JTOT}=1$ produces $n$p levels starting at $n=2$.

\subsubsection{Bound-state energies of Mg + NH at specified
magnetic fields}\label{testfiles:bound:MgNH}

\begin{tabular}{ll}
input file: & \file{bound-Mg\_NH.input}\\
executable: & \file{bound-Mg\_NH}\\
also required: & \file{data/pot-Mg\_NH.data}
\end{tabular}

\file{bound-Mg\_NH.input} locates the bound states of
MgNH at specified magnetic fields. It uses a plug-in basis-set suite for a
$^3\Sigma$ diatom colliding with a structureless atom. Radial potential
coefficients are obtained by RKHS interpolation of the potential points of
Sold\'an \etal~\cite{Soldan:MgNH:2009}. The coupled equations are solved using
the LDMD/AIRY hybrid propagation scheme.

The run locates a single bound state at four different magnetic fields from
370~G to 385~G, from which it may inferred that the state will cross threshold
near 387~G.


\subsubsection{Bound states of Mg + NH as a function of magnetic field}
\label{testfiles:field:MgNH}

\begin{tabular}{ll}
input file: & \file{field-Mg\_NH.input}\\
executable: & \file{field-Mg\_NH}\\
also required: & \file{data/pot-Mg\_NH.data}
\end{tabular}

\file{field-Mg\_NH.input} locates magnetic fields in the
range 0 to 400~G at which bound states exist for specific energies relative to
the lowest scattering threshold of Mg + NH in a magnetic field. It uses the
same basis-set suite and interaction potential as in section
\ref{testfiles:bound:MgNH}. The coupled equations are solved using the
LDMD/AIRY hybrid propagation scheme.

The run locates the same level as in section \ref{testfiles:bound:MgNH} at
energies of 0, 20 and 40 MHz $\times\ h$ below threshold, and shows that it
crosses threshold near 387.28~G.

\subsubsection{\texorpdfstring{Locating threshold crossings for $^{85}$Rb$_2$}{Locating threshold crossings for 85Rb2}}\label{basic:rb2:field}
\begin{tabular}{ll}
input file: & \file{field-basic\_Rb2.input}\\
executable: & \file{field-Rb2}
\end{tabular}

\file{field-basic\_Rb2.input} locates magnetic fields
where bound states cross the lowest scattering threshold for $^{85}$Rb$_2$.
These are the fields at which zero-energy Fesh\-bach resonances exist. It uses
a plug-in basis-set suite for a pair of alkali-metal atoms in a magnetic field,
including hyperfine interactions. It uses the potential of Strauss
\etal~\cite{Strauss:2010}, implemented with potential coefficients incorporated in
the executable.  The coupled equations are solved using the LDMD/AIRY hybrid
propagation scheme.

The basis-set suite for this interaction requires information about the
hyperfine properties of the atoms in an additional namelist block named
\namelist{\&BASIS9}. The potential
expansion comprises 3 terms: the singlet and triplet interaction potentials,
and the spin-spin dipolar term, which is modelled in the form
\begin{equation}
\lambda(R)=E_{\rm h}\alpha^2\left[\frac{g_S^2}{4(R/a_0)^3}+A\exp(-\beta R/a_0)\right].
\end{equation}

\subsubsection{\texorpdfstring{Bound states of $^{85}$Rb$_2$ as a function of
magnetic field}{Bound states of 85Rb2 as a function of
magnetic field}}\label{testfiles:field:85Rb2}

\begin{tabular}{ll}
input file: & \file{field-Rb2.input}\\
executable: & \file{field-Rb2}
\end{tabular}

\file{field-Rb2.input} locates bound states of
$^{85}$Rb$_2$ as a function of magnetic field, using the same potential and
basis-set suite as in section \ref{basic:rb2:field}. The calculation locates the
magnetic fields (in the range 750 to 850 G) at which bound states exist with
binding energies of 225, 175, 125, 75 and 25 MHz below the lowest threshold.
There are, however, two bound states that these calculations fail to find, as
they run almost parallel to the threshold, at about 140 and 220 MHz below it.
To locate these bound states, one would need to do a calculation using \BOUND.

\section{Program history}\label{history}

\BOUND\ was originally written by Jeremy Hutson in 1984 to calculate bound
states of van der Waals complexes by coupled-channel methods, using the same
structures as \MOLSCAT~\cite{molscat:2019} to generate the coupled equations.
Subsequent versions incorporated basis-set enhancements as they were made in
\MOLSCAT. A fundamental change was made in \BOUND\ version 5 (1993) to base the
convergence algorithm on individual eigenvalues of the log-derivative matching
matrix~\cite{Hutson:CPC:1994}, rather than its determinant. Versions 4 (1992)
and 5 (1993)~\cite{Hutson:bound:1993} were distributed via CCP6, the
Collaborative Computational Project on Heavy Particle Dynamics of the UK
Science and Engineering Research Council.

\BOUND\ was extended to handle calculations in external electric and magnetic
fields in 2007. \FIELD\ was written by Jeremy Hutson in 2010, using the same
structures as \BOUND\ to generate the coupled equations but designed to locate
bound states as a function of external field at fixed energy, rather than as a
function of energy.

There has been no fully documented publication of \BOUND\ since version 5, and
\FIELD\ has never been published.

\subsection{Principal changes in version \currentversion}\label{changes}

\begin{itemize}[leftmargin=13pt]
\item The basis-set plug-in mechanism has been extended to allow
    propagation in basis sets that are not eigenfunctions of the internal
    Hamiltonian $H_{\rm intl}$. This makes implementing new types of system
    much simpler than before, especially where the individual interaction
    partners have complicated Hamiltonians.

\item The basis-set plug-in functionality has been used to add new
    capabilities to carry out calculations in external fields (electric,
    magnetic, and/or photon) and to loop over (sets of) values of the
    fields.

\item The distance at which the calculation switches between short-range
    and long-range propagators ($R_{\rm mid}$) is now distinct from the
    distance at which the incoming and outgoing wavefunctions are matched
    ($R_{\rm match}$).

\item The programs now do an outwards propagation from $R_{\rm min}$ to
    $R_{\rm match}$ and an inwards propagation from $R_{\rm max}$ to
    $R_{\rm match}$. The node count is calculated without needing a third
    propagation from $R_{\rm match}$ to $R_{\rm min}$ or $R_{\rm max}$.

\item A more general mechanism for combining propagators has been
    implemented, allowing any sensible combination of propagators at short
    and long range.

\item A more general choice of log-derivative boundary conditions at the
    starting points for propagation is now allowed.

\item An additional propagation approach~\cite{MG:symplectic:1995} has been
    included, implemented by George McBane, which takes advantage of the
    symplectic nature of the multichannel radial Schr\"odinger equation.
\end{itemize}

\section{Acknowledgements}

We are grateful to an enormous number of people who have contributed routines,
ideas, and comments over the years. Any attempt to list them is bound to be
incomplete. We owe an enormous debt to the late Sheldon Green, who developed
the original \MOLSCAT\ program on which \BOUND\ and \FIELD\ are based. Robert
Johnson, David Manolopoulos, Millard Alexander and George McBane all
contributed propagation methods and routines. Alice Thornley developed routines
to propagate wavefunctions from log-derivative propagators. Maykel Leonardo
Gonz\'alez-Mart\'\i{}nez worked on the addition of structures for non-diagonal
Hamiltonians, including magnetic fields.

This work was supported by the U.K. Engineering and Physical Sciences Research
Council (EPSRC) Grant Nos.\ EP/P01058X/1, EP/P008275/1 and EP/N007085/1.




\bibliographystyle{elsarticle-num}
\bibliography{../all}







\end{document}